# Top-down proteomics on a microfluidic platform

Yuewen Zhang[1,‡], Maya A. Wright[1,‡], Pavan K. Challa[1,‡], Kadi-Liis Saar[1], Sean Devenish[1], Quentin Peter[1] & Tuomas P. J. Knowles[*,1,2]

[1] *Department of Chemistry, University of Cambridge, Lensfield Road, Cambridge, CB2 1EW, United Kingdom. E-mail: tpjk2@cam.ac.uk. Phone: +44 (0)1223 336344.*

[2] *Cavendish Laboratory, University of Cambridge, JJ Thomson Ave, Cambridge CB3 0HE, United Kingdom.*

**Protein identification and profiling is critical for the advancement of cell and molecular biology as well as medical diagnostics. Although mass spectrometry and protein microarrays are commonly used for protein identification, both methods require extensive experimental steps and long data analysis times. Here we present a microfluidic top down proteomics platform giving multidimensional read outs of the essential amino acids of proteins. We obtain hydro- dynamic radius and fluorescence signals relating to the content of tryptophans, tyrosines and lysines of proteins using a combination of diffusional sizing of proteins, label-free detection and on-chip labelling of proteins with a latent fluorophore in the solution phase. We thereby achieve identification of proteins on a single microfluidic chip by separating and mapping proteins in multidimensional space based on their characteristic physical parameters. Our results have significant implications in the development of easy and rapid platforms to use for native protein identification in clinical and laboratory settings.**

# 1 Introduction

Top down proteomic studies are of vital importance in adding to our fundamental knowledge of the human proteome as well as in the advancement of modern medicine. Indeed, proteomic studies on cancer cells and cerebrospinal fluid samples in Alzheimer's disease patients have provided precious insights into the mechanism of disease progression.[1–6] Currently, mass spectrometry is one of the most popular techniques available for top-down proteomics due to its high sensitivity, resolution, mass accuracy and dynamic range.[7, 8] Scientists also tried to combine tandem mass spectrometry with microfluidic method to identify proteins.[9] However, proteins are identified in the gas phase, typically by comparing protein mass spectra to predetermined sequence databases.[10–14] Another widely used method for identification is protein microarrays, which target proteins using analyte specific reagents such as antibodies, allowing quantitative information on the species present to be obtained.[15,16] It has been speculated that advances in protein identification studies could provide valuable information for the development of novel disease biomarkers, treatments in personalised medicine and targeted therapeutics, leading to the rise of a new generation of pharmaceuticals.[17–23] However, the currently available top down proteomic methods require extensive sample preparation steps, specialized mass spectrometry equipment and long experimental analysis times, and as of yet cannot be feasibly implemented in clinical settings.

Here, we propose a proof-of-principle technique which combines multidimensional data from microfluidic measurements on the physical parameters of proteins to achieve the resolution and sensitivity required for protein identification on a single microfluidic chip. We measure the protein hydrodynamic radius, detect label-free fluorescence intensity of tryptophan and tyrosine, as well as measure lysine residues that are labelled with a latent fluorophore. Each protein tested here has a unique combination of these four parameters, allowing us to separate and identify the proteins by mapping the ratio of the obtained parameters in multidimensional space. We thereby can identify those proteins tested here and demonstrate the potential that our novel microfluidic method has in the further development of top down proteomic assays for use in clinical diagnostic settings and in biophysical research laboratories for quick and easy protein identification.

## 2 Results and Discussion

**Microfluidic top-down proteomics measurements** The top-down proteomics device designed in this study consists of a diffusional sizing module and latent labelling module, allowing the protein hydrodynamic radius and characteristic fluorescence intensities from tryptophans, tyrosines and lysine to be determined in the solution phase in a single measurement.[23,24] The hydrodynamic radius of proteins and intrinsic fluorescence intensity for tryptophan and tyrosine are detected by our home-built microscope (Fig. 2) incorporate with deep UV-LED excitation at 280 nm. However, related fluorescence of lysine is detected by OPA-LED excitation at 365 nm light, which is because proteins were labelled with ortho-phthalaldehyde (OPA) fluorescence labelling dye to interact with primary amine group24. In order to calibrate the variations of fluorescence intensity of PDMS microfluidic devices under different daily experiments, we performed a calibration measurement where we separately imaged two different solutions (tryptophan and 4-methylumbelliferone (4MU)) of standard molecules of known concentrations with UV and OPA excitation wavelengths respectively (calibration channel, Fig. 2). After calibration, we used our intrinsic fluorescence set-up fitted with a tryptophan emission filter at 350 nm to size the native proteins by imaging the extent of their lateral diffusion into auxiliary buffer streams under steady laminar flow (sizing region, Fig. 2).[23] At the end of the diffusion channel we then imaged the characteristic tryptophan fluorescence (W detection region, Fig. 2). The emission filter was then changed to 305 nm in order to measure the related tyrosine fluorescence intensity for native proteins (Y detection region, Fig. 2). The tyrosine intensity was normalised by local background correction algorithm as previously described. Downstream of the sizing and the W detection regions, we used an on-chip latent labelling strategy described previously to conjugate the protein's lysine residues to OPA dye molecules.[24] The OPA labelling dye mixed with the protein via lateral diffusion. Thus, the characteristic fluorescence intensity from the lysine residues was imaged by switching the UV-LED light source at 280 nm wavelength to an OPA-LED light source at 365 nm wavelength.

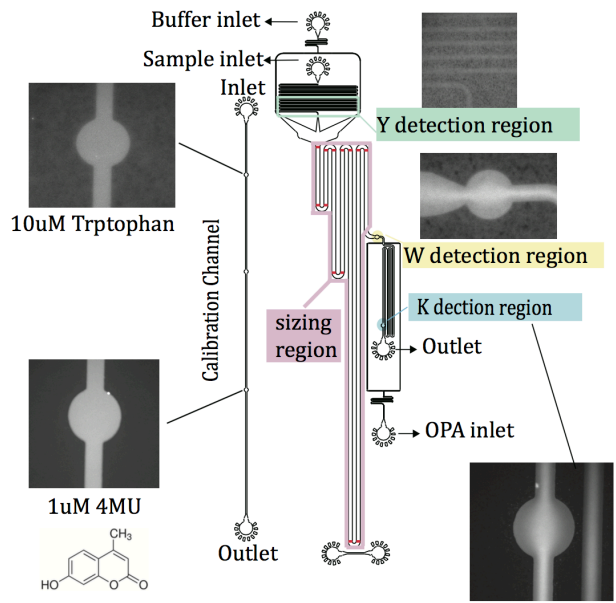

**Figure 1** The microfluidic top down identification device. The protein hydrodynamic radius is measured in the sizing region (pink) and the related intrinsic tryptophan (W) and tyrosine (Y) fluorescence intensities are imaged under UV excitation280 nm in the yellow and green regions, respectively. The light source is then switched to 365 nm excitation and the OPA fluorescence intensity coming from the protein's lysine residues (K) conjugated to OPA dye is imaged in the blue region. Before each experiment, standard dye molecules of known concentrations (tryptophan and 4MU) are imaged separately in the calibration channel to account for fluctuations of background fluorescence of PDMS. Because of the protein diffusion, both the W fluorescence intensity detection region and K detection region are not fully fluorescent at the detection region.

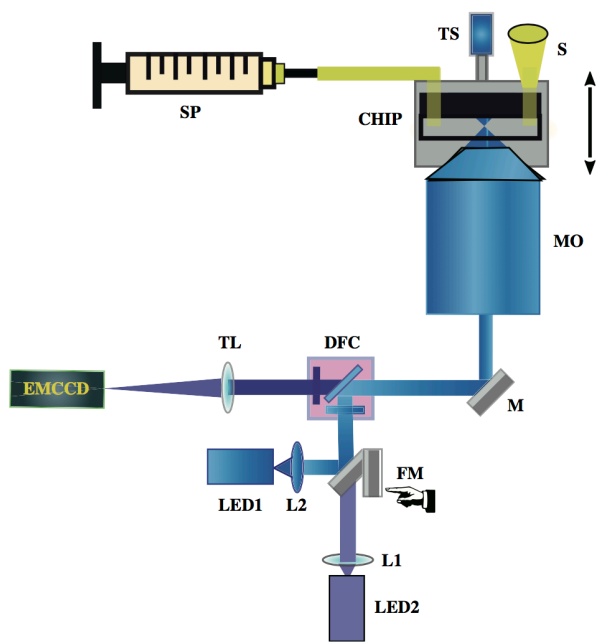

**Figure 2** Schematic representation of the home-built inverted fluorescence microscope used for visualisation of proteins in this work. The user is able to easily switch between two light sources of wavelength 280 nm and 365 nm using a flip mirror, and to change emission filters according to the amino acid residue under investigation.

**Protein identification in multidimensional space** In order to perform the multidimensional data analysis for protein identification, variations in sample concentration should also be considered. The imaged fluorescence intensities of the tryptophan, tyrosine and lysine residues were normalised for each repeat in each protein by protein absorbance measurements. Moreover, the three normalised fluorescence traces were converted into two fluorescence intensity ratios - $W_n/Y_n$ and $Y_n/K_n$ - to eliminate the concentration dependence. The protein concentration independent hydrodynamic radius was used to map the position of each of the seven proteins in multidimensional (Fig. 2).

For each of the seven areas along each of the axes the centre of the area was plotted as the average of the three repeat measurements ($\mu$) taken for the specific protein and the standard deviations ($\sigma$) along each of the three axes were calculated as the average of the seven individual standard deviations along that particular axis. Furthermore, the fourth test point obtained for each protein was randomly selected and the probability was calculated, then multiplied together to obtain the overall probability that the fourth test point would be identified as each protein in the database. Using this method, the heat map of the probabilities was plotted as shown in Fig. 3. However, it is hard to distinguish BSA and ovalbumin based on our database.

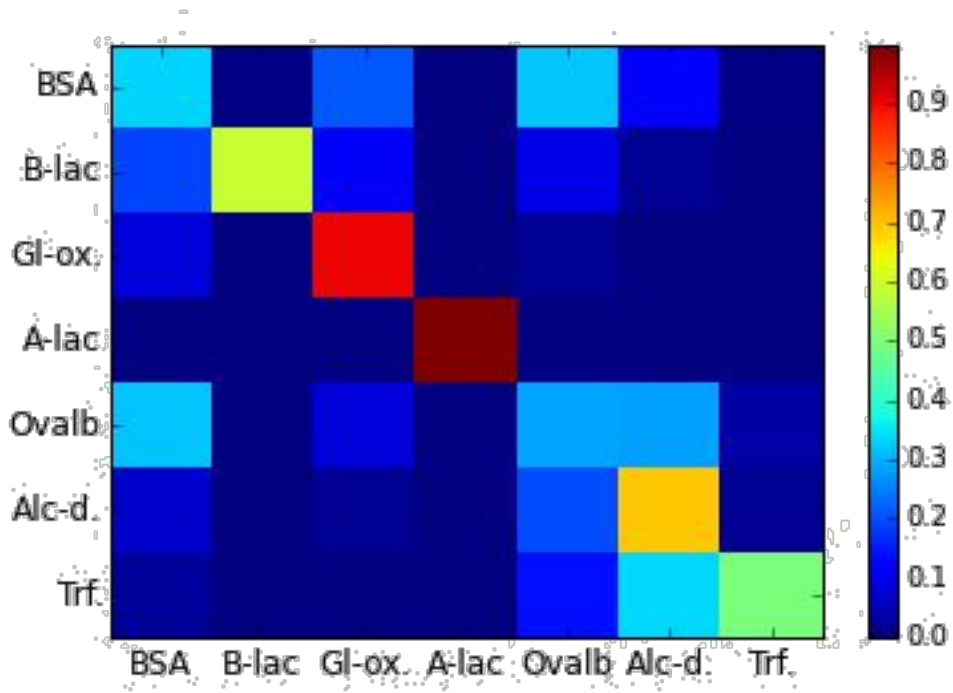

**Figure 3** The multidimensional data analysis for protein identification with top-down proteomics microfluidic device.

## 3 Conclusions

In summary, we are able to identify proteins of varying molecular weights and sequence in their native states by measuring their hydrodynamic radii and fluorescence intensities related to their tyrosine, tryptophan and lysine contents using a microfluidic top-down approach described in this work. Using the method described, we were able to obtain multidimensional information for a given protein in a single experiment. Because of its simplicity and low sample consumption, this method is also usable for detecting protein-protein interactions, protein stability and mobility, as well as available for clinical diagnostic assays.

**Methods**

**Preparation of samples and labelling dye** The proteins and their corresponding concentrations tested were: BSA (15µM and 30µM, Sigma Aldrich), β-lactoglobulin (β-lac) (353µM, Sigma Aldrich), glucose oxidase (30µM and 114µM), α-lactalbumin (α-lac) (231µM, Sigma Aldrich), ovalbumin (232µM), alcohol dehydrogenase (92µM and 140µM, Alfa Aesar), and human transferrin (100µM). All the proteins were prepared in the 25 mM phosphate buffer (pH 8).

Solutions for calibration at 280 nm and 365 nm were 10µM L-Tryptophan and 1µM 4MU, respectively, both in 400 mM potassium borate buffer (pH 9.7). These solutions were imaged separately on every chip in the detection regions of the calibration channels adjacent to the top-down identification microfluidic device.

The standard labelling solution was prepared with 12 mM OPA, 18 mM β-mercapto ethanol (BME) and 4% wt/vol sodium dodecyl sulfate (SDS) in 200 mM carbonate buffer, pH 10.5.

**UV-LED microscope** The schematic of optical layout is shown in Fig. 2. Light from a 280 nm LED (Thorlabs M280L3) and 365 nm LED (Thorlabs M365L2) are individually selected depending upon the experiment by a flip mirror. The light is passed through an aspherical lens of focal length 20 mm to get a nearly collimated output beam. The beam is then incident on a filter cube, which consists of an excitation filter, dichroic mirror and an emission filter. Light reflected by the dichroic mirror is focused onto the sample flowing in a microfluidic chip by an infinity corrected UV objective lens (Thorlabs LMU-10X-UVB) of numerical aperture NA=0.25. The emitted fluorescent light from the sample is collected through the same objective, an emission filter, and thereafter focused onto a EMCCD camera (Rolera EM-C2) by a tube lens of focal length 200 mm. All the optics used in the set-up are made out of fused silica for high transmission in the UV region.

**Microfluidic device fabrication and experiments** Microfluidic devices were cast using polydimethylsiloxane (PDMS ) (Sylgard 184 kit, Dow Corning) from a silicon wafer master imprinted with the device channels 50μm high based on standard soft-lithography techniques[25]. Carbon black nano-powder (Sigma-Aldrich) was added to the PMDS before curing to create black devices, thus minimising unwanted autofluorescence from PDMS under UV illumination during the measurements. Devices were bonded to a quartz slide (Alfa Aesar, 76.2x25.4x1.0mm) using a plasma treatment (Electronic Diener Femto plasma bonder). The channels were then filled from the outlet with phosphate buffer using a glass syringe (Hamilton, 500μL), needle (Neolus Terumo, 25 gauge, 0.5 x 16mm), and polyethene tubing (Scientific Laboratory Supplies, inner diameter 0.38mm, outer diameter 1.09mm).

The sample, buffer and labelling solution were loaded in their respective device inlets using gel loading pipette tips. The calibration channel was filled in the same way and the standard UV calibration solution was loaded in a gel pipette tip. Fluid flow through the channels was controlled using neMESYS syringe pumps (Cetoni GmbH) at a flow rate of 200 μLh−1 in the top-down identification channels and 80 μLh−1 in the calibration channel.

Once the flows stabilised, hydrodynamic radii of proteins were measured and fluorescence intensities of tryptophan and tyrosine were detected using our UV-LED set-up fitted with the appropriate emission filter (Tryptophan: 350 nm; Tyrosine: 305 nm), and excitation wavelength of 280 nm. In order to get the fluorescence intensity related to the protein's lysine content, OPA dye solution was mixed with the protein on-chip via lateral diffusion from both sides, with a mixing time of at least 3 s at a flow rate of 200 μL h$^{-1}$ to ensure fully labelling. The OPA detection region in the top-down channels and the calibration channels as well as the background were then imaged using our set-up fitted with a 365 nm LED and corresponding dichroic filter set. The protein fluorescence intensities were calculated from the acquired images using ImageJ (see details in Supporting information).

**Data Analysis** The hydrodynamic radii of the proteins were calculated based on 12 images along the diffusion channel. The 12 images were processed to produce a set of 2D diffusion profiles, which were then fitted to a set of simulated basis functions (Fig. S). These basis functions contain information both on the spatial diffusion (along the channel width) and temporal diffusion (along the channel length), which allows the sample diffusion coefficients to be determined. A maximal entropy basin hopping algorithm with a Broyden-Fletcher-Goldfarb-Shannon minimization was used to fit a linear combination of the simulated basis functions to the experimental diffusion profile, yielding the average radius of the analyte[23].

Based on our approach described, multi-information of proteins can be determined. To perform the multidimensional data analysis, we first normalised the fluorescence intensities for each repeat in each protein by protein absorbance using the following equations:

$$W_n = \frac{W}{W_{cal} \cdot A_{280}}$$

$$K_n = \frac{K}{K_{cal} \cdot A_{280}} \quad (1)$$

$$Y_n = \frac{Y}{A_{280}}$$

where $W_n$, $K_n$ and $Y_n$ are the normalised tryptophan, lysine or tyrosine intensities, W, K and Y are the tryptophan, lysine and tyrosine intensities obtained in the microfluidic measurement, $W_{cal}$ and $K_{cal}$ are the intensities of the standard UV and OPA calibration solutions for background correction, and A280 is the absorbance of the protein.

We then calculated the average (µ) and standard deviation (σ) of each parameter ($W_n$, $K_n$, $Y_n$ and $R_h$) across three repeats for each protein to define the 3D space. In order to eliminate the concentration dependence the three fluorescence traces were converted into two fluorescence intensity ratios - $K_n/W_n$ and $K_n/Y_n$.

The fourth data point was randomly selected and used as a test point for protein identification. In order to identify the test proteins within the multidimensional space were modelled as standard normal distributions N (µ, σ) with µ and σ determined as previously described. Such distributions were obtained for all the seven proteins along all the axes. The z-scores were then calculated under the assumption of each of the test points being each of the seven proteins. All the z-scores were converted into the probabilities given the specific normal distribution and the three probabilities were multiplied and converted to logarithm. Then, the products of the probabilities were plotted in the Fig. 2, y-axis and as such the probabilities are normalised along each row.

**Correspondence** Correspondence and requests for materials should be addressed to Tuomas P.J. Knowles (email: tpjk2@cam.ac.uk).

**Figure 1** The microfluidic top down identification device. The protein hydrodynamic radius is measured in the sizing region (pink) and the related intrinsic tryptophan (W) and tyrosine (Y) fluorescence intensities are imaged under UV excitation280 nm in the yellow and green regions, respectively. The light source is then switched to 365 nm excitation and the OPA fluorescence intensity coming from the protein's lysine residues (K) conjugated to OPA dye is imaged in the blue region. Before each experiment, standard dye molecules of known concentrations (tryptophan and 4MU) are imaged separately in the calibration channel to account for fluctuations of background fluorescence of PDMS. Because of the protein diffusion, both the W fluorescence intensity detection region and K detection region are not fully fluorescent at the detection region.

**Figure 2** Schematic representation of the home-built inverted fluorescence microscope used for visualisation of proteins in this work. The user is able to easily switch between two light sources of wavelength 280 nm and 365 nm using a flip mirror, and to change emission filters according to the amino acid residue under investigation.

**Figure 3** The multidimensional data analysis for protein identification with top-down proteomics microfluidic device.